\documentclass[journal,final,twocolumn,twoside]{IEEEtran}
\usepackage{amssymb}
\usepackage{amsfonts}
\usepackage{balance}

\usepackage{tikz}
\usepackage{cite}
\usepackage[cmex10]{amsmath}
\usepackage{url}
\usepackage{multirow}
\allowdisplaybreaks[1]

\begin{document}

\title{Efficient Majority-Logic Decoding of Short-Length Reed--Muller Codes at Information Positions}

\author{Peter~Hauck, Michael~Huber,~\IEEEmembership{Member,~IEEE,} Juliane~Bertram, Dennis~Brauchle and~Sebastian~Ziesche%
\thanks{Manuscript received July 4, 2011; revised June 20, 2012.
The work of P.~Hauck was supported by Proyecto MTM2010-19938-C03-02, Ministerio de Ciencia e Innovaci\'{o}n.
The work of M.~Huber was supported by the Deutsche Forschungsgemeinschaft (DFG) via a Heisenberg grant (Hu954/4) and a Heinz Maier-Leibnitz Prize grant (Hu954/5).}
\thanks{The authors are with the Wilhelm Schickard Institute for Computer Science, Eberhard Karls Universit\"at T\"ubingen, Sand~13,
D-72076 T\"ubingen, Germany (corresponding authors' e-mail addresses: \{hauck,\,huber\}@informatik.uni-tuebingen.de).}}%

%\markboth{IEEE Transactions on Communications (submitted)}%,~Vol.~6, No.~1, January~2012}%
%{Hauck and Huber \MakeLowercase{\textit{et al.}}: Efficient Majority-Logic Decoding of RM Codes at Information Positions}

\newtheorem{theorem}{Theorem}[section]
\newtheorem{corollary}[theorem]{Corollary}
\newtheorem{lemma}[theorem]{Lemma}
\newtheorem{proposition}[theorem]{Proposition}
\newtheorem{problem}[theorem]{Problem}
\newtheorem{example}[theorem]{Example}
\newtheorem{examples}[theorem]{Examples}
\newtheorem{decoding}[theorem]{Chen's decoding procedure}
\newtheorem{notation}[theorem]{Notation}
\newtheorem{remark}[theorem]{Remark}

\numberwithin{equation}{section}

\maketitle

\begin{abstract}
Short-length Reed--Muller codes under majority-logic decoding are of particular importance for efficient hardware implementations in real-time and embedded systems. This paper significantly improves Chen's two-step majority-logic decoding method for binary Reed--Muller codes $\text{RM}(r,m)$, $r \leq m/2$, if --- systematic encoding assumed --- only errors at information positions are to be corrected.  Some general results on the minimal number of majority gates are presented that are particularly good for short codes. Specifically, with its importance in applications as a 3-error-correcting, self-dual code, the smallest non-trivial example, $\text{RM}(2,5)$ of dimension 16 and length 32, is investigated in detail. Further, the decoding complexity of our procedure is compared with that of Chen's decoding algorithm for various Reed--Muller codes up to length $2^{10}$.
\end{abstract}

\begin{IEEEkeywords}
Majority-logic decoding, Reed--Muller codes, information positions,  real-time and embedded systems.
\end{IEEEkeywords}

\normalsize

\section{Introduction}\label{sect_intro}
\IEEEPARstart{M}{ajority-logic} decoding is a simple procedure, particularly suited for hardware implementations  using most elementary circuits. Reed--Muller codes constitute the most prominent examples for which majority-logic decoding is possible. In 1971, Chen~\cite{Chen71} showed that many finite geometry codes, among them all  Reed--Muller codes $\text{RM}(r,m)$, $r\leq m/2$, allow a two-step majority-logic decoding, improving significantly the decoding complexity in comparison with the well-known step-by-step decoding algorithm of Reed~\cite{Reed54}. Although recursive soft-decision decoding algorithms for Reed--Muller codes~\cite{SchnBos95,Dumer04} have complexity advantages especially for long block lengths, majority-logic decoding remains the premier choice for simple hardware implementation issues.
In particular, the study of Reed--Muller codes of short block lengths (up to $2^{10}$ bits) under majority-logic decoding is of special interest in applications where there is little or no tolerance for delay on system response times, i.e., in hard real-time and embedded systems such as in airbag systems, anti-lock braking systems (ABS) or medical systems like heart pacemakers. These systems usually interact at a low level with physical hardware. In such cases, short-length Reed--Muller codes may be applied as stand-alone codes. In addition, these codes may be employed as component codes in concatenated coding schemes. While cyclic redundancy-check (CRC) codes have commonly been used in real-time and embedded environments focusing only on error detection, the enduring trend towards ever smaller and complex electronic components rises further reliability issues with an increasing demand for error correction. One example is the use of extended single-error-correcting Hamming co
 des in real-time control of safety-critical applications in the automotive and industrial markets~\cite{Freescale,Renesas}. Another important application area concerns in-built error correction in memory devices, for instance, in flash memories~\cite{Huang11} and static random-access memories (SRAMs)~\cite{Issi11}.

In the following, we restrict ourselves to binary Reed--Muller codes $\text{RM}(r,m)$, $r\leq m/2$. If errors at all positions of a word are to be corrected, assuming that at most $2^{m-r-1}-1$ errors occurred, it is easy to see that an optimal implementation of Chen's procedure requires $2^{m-r}(2^{m-r}-2)$ majority gates in the first step (Proposition~\ref{prop_inters}) and $2^m$ majority gates in the second step, namely one for each position. The aim of this paper is to draw attention to the fact that the decoding complexity can be significantly reduced if under systematic encoding only errors at information positions are to be corrected, assuming that at most $t=2^{m-r-1}-1$ errors occurred in the \emph{whole} word. Since correctness of information bits is all what is needed, this approach is justified and it results in the advantage of providing more efficient hardware implementations in the above mentioned fields of applications.

If solely errors at information positions are to be corrected, only $k$ majority gates are needed for the second step where $k= \sum_{i=0}^r \binom{m}{i}$ is the dimension of $\text{RM}(r,m)$. However, it is a complicated combinatorial problem to determine the minimal number of majority gates for the first decoding step. Simple cases like $\text{RM}(1,m)$  and $\text{RM}(2,4)$ can be solved quickly (Subsections~\ref{subsect_RM1_r} and~\ref{subsect_RM2_4}). In the general case, upper bounds are presented (Proposition~\ref{prop_sect3})  that are particularly good for short codes and a non-trivial lower bound can be obtained from a solution to an integer linear programming problem (Proposition~\ref{prop3.4}), but to obtain the exact number appears to be difficult. With its importance as a $3$-error-correcting, self-dual code of length $32$ and dimension $16$, we investigate $\text{RM}(2,5)$ thoroughly in Subsection~\ref{subsect_RM2_5}. Since the  minimal number of majority gate
 s for the first step might be dependent on the choices of the information positions, we calculate first the number of orbits of the automorphism group $\text{AGL}(5,2)$ of  $\text{RM}(2,5)$ on the collection of $16$-sets that are information positions. It turns out that there are $7$ orbits and we list representatives of each orbit in Proposition~\ref{prop_sect4}. For each of them there is a solution with $30$ majority gates and we show that independent of the chosen information positions, $30$ is in fact the minimal number (Theorem~\ref{minnumb_gates}). In conclusion, $30$ gates mean a significant reduction of the $48$ majority gates that are required for the first step if all positions are to be corrected. In Section~\ref{comparison}, we compare the decoding complexity of our procedure with that of Chen's algorithm for various Reed--Muller codes up to length $2^{10}$. Concluding remarks are presented in Section~\ref{conclusion}.

\section{Two-step majority-logic decoding of Reed--Muller codes}\label{sect_2step}

The binary Reed--Muller code $\text{RM}(r,m)$ has length $2^m$, dimension $k= \sum_{i=0}^r \binom{m}{i}$ and minimal distance $2^{m-r}$. There are many ways to describe Reed--Muller codes (cf., e.g.,~\cite[Ch.\,1, Sect.\,13 \mbox{and} Ch.\,16, Sect.\,3]{HB98} or~\cite[Ch.\,13]{Blahut03}). For our purposes, the geometric one is most suitable: By a $d$-flat in $\mathbb{Z}_2^m$ we mean an affine subspace of dimension $d$, i.e. a coset of a linear subspace of dimension $d$ in $\mathbb{Z}_2^m$. We identify the positions of a vector in $\mathbb{Z}_2^{2^m}$ (written as a row vector) by the vectors $v_0,\ldots, v_{2^m-1}$ in $\mathbb{Z}_2^m$ given in some ordering. Then $\text{RM}(r,m)$ is generated as a subspace of $\mathbb{Z}_2^{2^m}$ by all characteristic vectors $\chi_U$ of $d$-flats $U$ in $\mathbb{Z}_2^m$ with $d \geq m-r$~\cite[Ch.\,16, Sect.\,3.2]{HB98}. (Recall that the $i$-th component of $\chi_U$ is $1$ if and only if $v_i \in U$.)

Admittedly, the notation used here is somewhat sloppy: Given $r$ and $m$, we denote by $\text{RM}(r,m)$ actually not one code but a family of equivalent codes depending on the chosen ordering in $\mathbb{Z}_2^m$. Usually, the notation  $\text{RM}(r,m)$ is reserved for the code which corresponds to the lexicographic ordering of the row vectors in  $\mathbb{Z}_2^m$. Since the results in this paper are independent of the type of ordering, the ambiguous meaning of $\text{RM}(r,m)$ is not harmful and is justified for the sake of notational simplicity.

The following lemma is the basis for Chen's two-step majority-logic decoding procedure:

\begin{lemma}\label{lem_flats}
Let $r,m$ be non-negative integers, $r<m$.

\begin{enumerate}

\item[(a)] Given an $r$-flat U of $\mathbb{Z}_2^m$, there are exactly $2^{m-r}-1$ $(r+1)$-flats in $\mathbb{Z}_2^m$ that contain $U$.

\item[(b)] Given $v \in \mathbb{Z}_2^m$, there exist at least $2^{m-r}-2$ \mbox{$r$-flats} in $\mathbb{Z}_2^m$ that intersect pairwise in $v$.
(In fact, it can be shown that actually at least $2^{m-r}+1$ such $r$-flats exist, but this is not needed in the sequel.)
\end{enumerate}

\end{lemma}

\begin{IEEEproof}
(a) is obvious.

(b): This follows from~\cite[Lemmas~2 and~6]{Chen71}; see also~\cite[Sects.~II and~III]{Chen72}.
\end{IEEEproof}

\begin{decoding}[cf.~\cite{Chen71}]\label{chen}
We assume that $m \geq 3$ and $1\leq r \leq m/2$. $\text{RM}(r,m)$ is $t$-error correcting with $t=2^{m-r-1}-1$. Under the assumption that at most $t$ errors have occurred, the following procedure corrects these errors:
Let $y \in \mathbb{Z}_2^{2^m}$ be the received word, $j \in \{0, \ldots, 2^m-1\}$ a position. By Lemma~\ref{lem_flats}~(b), choose $2^{m-r}-2$ $r$-flats $U_1,\ldots,U_{2^{m-r}-2}$ in $\mathbb{Z}_2^m$ such that $U_s \cap U_{s^\prime} =\{v_j\}$ for $s \neq s^\prime$. For each $U_s$, choose $2^{m-r}-2$ $(r+1)$-flats $V_{s,1}, \ldots, V_{s,2^{m-r}-2}$ that contain $U_s$ (Lemma~\ref{lem_flats}~(a)). Note that $\chi_{V_{s,i}} \in \text{RM}(m-r-1,m)=\text{RM}(r,m)^\perp$.

\textit{Step~1.\,} Compute $y \cdot \chi_{V_{s,i}}$ for $i=1, \ldots, 2^{m-r}-2$. If at least $2^{m-r-1}$ of these values are $1$, then $y$ contains an odd number of errors at the positions of $U_s$, otherwise it contains an even number of errors. Perform this step for $s=1, \ldots,2^{m-r}-2$.

\textit{Step~2.\,} If at least for $2^{m-r-1}$ of the $U_s$ the number of errors in $y$ at the positions of $U_s$ is odd, then position $j$ is not correct, otherwise correct.
\end{decoding}

To correct all positions, clearly $2^m$ majority gates (each with $2^{m-r}-2$ inputs) are needed in the second step of~\ref{chen}. The number of majority gates needed in total for the first step depends on how many  of the $r$-flats are required for the correction of all positions. Every $r$-flat could in principle be used for all $2^r$ positions $v_j$ that it contains. This leads to the question if one can choose $r$-flats in such a way that any two of them intersect in at most one $v_j$ and such that every $v_j$ is contained in exactly $2^{m-r}-2$ of these $r$-flats. In this case, a totality of $2^m(2^{m-r}-2)/2^r=2^{m-r}(2^{m-r}-2)$ $r$-flats would suffice for the first step and hence the same number  of majority gates (again each with $2^{m-r}-2$ inputs). This is actually possible:

\begin{proposition}\label{prop_inters}
There exist $2^{m-r}(2^{m-r}-2)$ \mbox{$r$-flats} in $\mathbb{Z}_2^m$ such that the intersection of any two of them has at most size $1$ and every $v \in \mathbb{Z}_2^m$ is contained in exactly $2^{m-r}-2$ of these $r$-flats.
\end{proposition}

\begin{IEEEproof}
Choose $2^{m-r}-2$ linear subspaces of dimension $r$ in $\mathbb{Z}_2^m$, $U_1,\ldots,U_{2^{m-r}-2}$, such that $U_s \cap U_{s^\prime}= \{0\}$ for all $s\neq s^\prime$ (Lemma~\ref{lem_flats}~(a)). Choose linear subspaces $W_1, \ldots,W_{2^{m-r}-2}$ such that $\mathbb{Z}_2^m=U_p \oplus W_p$, $p=1,\ldots,2^{m-r}-2$. Clearly, $\left| W_p\right|=2^{m-r}$. Let $W_p =\{w_{p,1},\ldots,w_{p,2^{m-r}}\}$ and set $U_{p,q}=w_{p,q}+U_p$, $q=1,\ldots,2^{m-r}$, $p=1,\ldots,2^{m-r}-2$. Then for every $v \in \mathbb{Z}_2^m$ and for every $p \in \{1,\ldots,2^{m-r}-2\}$ there exists exactly one $q=q(v,p)$ and exactly one $u_p \in U_p$ with $v=w_{p,q}+u_p$. It follows that the family $U_{p,q}$ of $r$-flats fulfills the desired requirements.
\end{IEEEproof}

\section{Error correction at information positions}\label{sect_infopos}

As in Section~\ref{sect_2step}, we assume that $m \geq 3$ and $1 \leq r \leq m/2$. Furthermore we suppose that a systematic encoding procedure is used for $\text{RM}(r,m)$, i.e. an information vector $y$ of length $k$, $k=\sum_{i=0}^r \binom{m}{i}$ being the dimension of $\text{RM}(r,m)$, is encoded in such a way that $y$ appears unchanged at $k$ prescribed positions of the codeword, the information positions. This can be achieved by multiplying $y$ with a generator matrix $\widetilde{G}$ of $\text{RM}(r,m)$ in systematic form with respect to the chosen information positions, i.e. the columns of $\widetilde{G}$ at the information positions constitute the $k \times k$ unit matrix.

Note that we use here a wider concept of systematic encoding (in accordance with~\cite[Ch.\,1, p.\,16]{HB98}) than the one frequently encountered where it is assumed that the information vector $y$ appears at consecutive positions at the beginning (or at the end) of the corresponding codeword (see, e.g.,~\cite[Def.\,3.2.4]{Blahut03}). In the extended meaning used here, every linear code can be encoded systematically and usually in many different ways (cf. the case of $\text{RM}(2,5)$ in Section~\ref{specialcases} of this paper).

It is our aim in this section to investigate how many majority gates in the first step of Chen's decoding procedure are needed if only errors at the information positions are to be corrected. In other words, we want to determine the minimal number of $r$-flats in $\mathbb{Z}_2^m$ such that each $v\in \mathbb{Z}_2^m$ corresponding to an information position is contained in $2^{m-r}-2$ of these $r$-flats and where any two of these intersect in $v$. This number may depend on the chosen information positions. We emphasize that the correction of errors at information positions assumes that at most $t = 2^{m-r-1}-1$ errors occurred in the whole codeword. Whether this assumption is valid can be checked by assigning to the vector of information positions (after correction) the codeword with respect to the chosen systematic encoding procedure. If this codeword has Hamming distance at most $t$ from the received word, the assumption is justified.

We denote the positions of vectors in $\mathbb{Z}_2^{2^m}$ by $0,\ldots,2^m-1$ corresponding to an ordering $v_0,\ldots,v_{2^m-1}$ of the vectors in $\mathbb{Z}_2^m$.

We fix the following notation:

\begin{notation}\label{nota}

\begin{enumerate}

\item[(a)] $I(r,m)=\{J \subseteq \{0,\ldots,2^m-1\}$: $\left|J\right|=k$, the columns at positions $j \in J$ of a generator matrix of $\text{RM}(r,m)$ are linearly independent\}.

$I(r,m)$ consists of all possible sets of information positions for $\text{RM}(r,m)$; note that $I(r,m)$ is clearly independent of the chosen generator matrix. It is clear that for any $J \in I(r,m)$ and any generator matrix $G$ of $\text{RM}(r,m)$, $G$ can be transformed to a generator matrix in standard form with respect to the positions in $J$.
As for any linear code, the following holds for $\text{RM}(r,m)$.
Let $J\subseteq \{0,\ldots,2^m-1\}$, $\left|J\right|=k$. Then:
\begin{align*}{}
    &J \in I(r,m)  \\
     \Longleftrightarrow \; &\text{supp}(c) \nsubseteq \{0,\ldots,2^m-1\}\setminus J \;\, \\
    &\text{for all}  \;\, 0 \neq c \in\text{RM}(r,m)\\
    \Longleftrightarrow \; &\text{supp}(\tilde{c}) \nsubseteq J\;\, \\ 
    &\text{for all} \;\, 0 \neq \tilde{c} \in\text{RM}(r,m)^\perp=\text{RM}(m-r-1,m).
\end{align*}

\item[(b)] A family $\mathcal{F}=\{U_1,\ldots,U_l\}$ of $r$-flats in $\mathbb{Z}_2^m$ is called \emph{admissible} with respect to $J \in I(r,m)$ (in short, \mbox{$J$-admissible}) if the following holds:
For every $j \in J$ there exist $2^{m-r}-2$ $r$-flats $U_{i_1},\ldots,U_{i_{2^{m-r}-2}}$ in $\mathcal{F}$ such that $U_{i_p} \cap U_{i_q}=\{v_j\}$ for all \mbox{$p,q \in \{1, \ldots, 2^{m-r}-2\}$}, $p \neq q$. We say then that $U_{i_1},\ldots,U_{i_{2^{m-r}-2}}$ are \emph{used at position $j$} and that position $j$ is \emph{active} in $U_{i_1},\ldots,U_{i_{2^{m-r}-2}}$.

\item[(c)] For $J \in I(r,m)$ we denote by $\text{min}_J(r,m)$ the minimal cardinality of an $J$-admissible family of $r$-flats.

\item[(d)] $\mu(r,m)=\text{min}\{\text{min}_J(r,m): J \in I(r,m)\}$.

Hence $\mu(r,m)$ is the minimal number of majority gates in the first step of Chen's decoding procedure needed to correct all errors at suitably chosen information positions (if at most $t=2^{m-r-1}-1$ errors occurred in the whole word).

\end{enumerate}

\end{notation}

By Proposition~\ref{prop_inters}, we have
\begin{equation}\label{E1}
\mu(r,m) \leq 2^{m-r}(2^{m-r}-2).
\end{equation}

Moreover,

\begin{equation}\label{E2}
\bigg\lceil\frac{k \cdot (2^{m-r}-2)}{2^r}\bigg\rceil \leq \mu(r,m) \leq k \cdot(2^{m-r}-2).
\end{equation}

The upper bound in~(\ref{E2}) is trivial. The lower bound would be achieved if in an admissible family of $r$-flats all $r$-flats consist only of vectors corresponding to information positions and every such $r$-flat can be used at each of the positions corresponding to the vectors it contains; in particular, any two members of the family have at most one vector in common.
The upper bound in~(\ref{E1}) is superior to the one in~(\ref{E2}) if $r$ is large (approximately, $r\geq m/4$).

Both upper bounds can be slightly improved as the next proposition shows, and with more care further improvements are possible. However, it seems to be difficult to obtain sharp upper bounds or even the exact value of $\mu(r,m)$.
The statement of Proposition~\ref{prop_sect3} is derived by using a special set $J$ of information positions and part~(a) of~\ref{prop_sect3} by constructing a particular family of $r$-flats covering the vectors in $\mathbb{Z}_2^m$ corresponding to the positions in $J$. 

\begin{proposition}\label{prop_sect3}
Let $1\leq r \leq m/2$, $m\geq 3$.
\begin{enumerate}

\item[(a)] 
$\mu(r,m) \leq k \cdot (2^{m-r}-3) + \binom{m-b}{r-b} + \left\lceil\frac{1}{r+1} \cdot \sum_{s=0}^{b-1} \binom{m-1-s}{r-s} \right\rceil$
where $b=\lceil \log_2(r+1)\rceil$.
(Note $\binom{m-b}{r-b} + \left\lceil\frac{1}{r+1} \cdot \sum_{s=0}^{b-1} \binom{m-1-s}{r-s} \right\rceil < k= \sum_{s=0}^{r} \binom{m}{s}$.)

\item[(b)] $\mu(r,m) \leq 2^{m-r} (2^{m-r}-2) - \left\lfloor\frac{m}{r} \right\rfloor \cdot \sum_{s=0}^{m-2r-1} \binom{m-r}{s}$.
\end{enumerate}
\end{proposition}

\begin{IEEEproof}
The proof is given in the Appendix.
\end{IEEEproof}

We now describe a method to improve (in principle) the lower bound for $\mu(r,m)$ given in~(\ref{E2}).

Let $J$ be any set of information positions for $\text{RM}(r,m)$ and let $\mathcal{F}$ be a family of $J$-admissible $r$-flats. 
Let $x_i$ be the number of $r$-flats in $\mathcal{F}$ that are used for exactly $i$ positions in $J$, $i=1,\ldots,2^r$ (that is, such an $r$-flat contains exactly $i$ vectors $v_{j_1}, \ldots,v_{j_i}$ with $j_1,\ldots, j_i \in J$ such that the $r$-flat is used at positions $j_1,\ldots, j_i$; note that such an $r$-flat could also contain further vectors $v_j$ with $j \in J$, that are not used at position $j$).
Then $\sum_{i=1}^{2^r} x_i=\left|\mathcal{F}\right|$.
We show that
\begin{align}{}
& x_i \geq 0\;\,\mbox{for}\;\, i=1,\ldots,2^r \label{E4}\\
& \sum_{i=1}^{2^r} i\cdot x_i \geq k\cdot (2^{m-r}-2)\label{E5}\\
& \sum_{i=2}^{2^r} \binom{i}{2} x_i \leq \binom{k}{2}\label{E6}.
\end{align}
(\ref{E4}) and (\ref{E5}) are clear. It remains to confirm~(\ref{E6}).
Let $X,Y$ be two arbitrary distinct $r$-flats in $\mathcal{F}$ such that $X$ is used at positions $j_1,\ldots,j_h \in J$ and $Y$ at positions $k_1, \ldots , k_g \in J$. Then $\left| \{j_1,\ldots,j_h\} \cap \{k_1,\ldots,k_g\}\right| \leq 1$ since the $r$-flats used at position $j \in J$ intersect pairwise in $v_j$. Therefore, the $\binom{h}{2}$ $2$-subsets of $\{j_1,\ldots,j_h\}$ are pairwise distinct from the $\binom{g}{2}$ $2$-subsets of $\{k_1,\ldots,k_g\}$. Since $\left|J\right|=k$, it follows that $\sum_{i=2}^{2^r} \binom{i}{2} x_i \leq \binom{k}{2}$.

These conditions yield immediately

\begin{proposition}\label{prop3.4}
Let $1 \leq r \leq m/2$, $k=\sum_{i=0}^r \binom{m}{i}$. Consider the following integer linear programming problem (ILP):
Minimize $F(x_1,\ldots,x_{2^r})=\sum_{i=1}^{2^r} x_i$, $x_i \in \mathbb{Z}$, with respect to conditions~(\ref{E4}),~(\ref{E5}),~(\ref{E6}). 
If $F_{\text{min}}$ is the minimal value of $F(x_1,\ldots,x_{2^r})$ for this ILP, then $F_{\text{min}} \leq \mu(r,m)$.
\end{proposition}

\begin{table*}[!t] %[!t] for top table* for two-columns floats
\renewcommand{\arraystretch}{1.3}
\caption{Some upper and lower bounds for $\mu(r,m)$.}\label{Table1}

\begin{center}
\begin{tabular}{|c|c||cc|cc|}
  \hline
   $r$ & $m$ & \multicolumn{2}{|c|}{lower bound for $\mu(r,m)$} & \multicolumn{2}{|c|}{upper bound for $\mu(r,m)$}\\
  \hline \hline
  1 &  3        & 4 & (\ref{E2}) & 5 & (\ref{prop_sect3}~(b))  \\
  1 & $\geq 4$  & $\frac{(m+1)(2^m-m-4)}{2}$ & (\ref{prop3.4}) & $\left\lceil \frac{(m+1)(2^m-5)}{2}\right\rceil$ & (\ref{prop_sect3}~(a))    \\
  2 & 4         & 6 & (\ref{E2}) \& (\ref{prop3.4}) & 8 & (\ref{E1}) \& (\ref{prop_sect3}~(b))  \\
  2 & 5         & 28 & (\ref{prop3.4}) & 46 & (\ref{prop_sect3}~(b))  \\
  2 & 6         & 129 & (\ref{prop3.4}) & 209 & (\ref{prop_sect3}~(b))  \\
  2 & 7         & 464 & (\ref{prop3.4}) & 849 & (\ref{prop_sect3}~(a))  \\
  3 & 6         & 33 & (\ref{prop3.4}) & 48 & (\ref{E1}) \& (\ref{prop_sect3}~(b)) \\
  3 & 7         & 165 & (\ref{prop3.4}) & 222 & (\ref{prop_sect3}~(b))  \\
  \hline
\end{tabular}
\end{center}
\end{table*}

We collect some upper and lower bounds for $\mu(r,m)$ for various small values of $r$ (and $m$) in Table~\ref{Table1}. We indicate which result was used to obtain the best bounds.

\section{Some special cases}\label{specialcases}

We show now that for $r=1$ the lower bounds in Table~\ref{Table1} are in fact the true values for $\mu(1,m)$. We also prove that
$\mu(2,4)$ is in fact $7$. Because of its importance as a $3$-error-correcting, self-dual code of length $32$ and dimension $16$, we investigate $\text{RM}(2,5)$ more thoroughly in Subsection~\ref{subsect_RM2_5} and show that $\mu(2,5)=30$, hence significantly closer to the lower bound $28$ given above than to the upper bound $46$.

\subsection{$\text{RM}(1,r)$}\label{subsect_RM1_r}

\begin{enumerate}

\item[(a)] $\mu(1,3)=4$:\\
Here, $k=4$. W.l.o.g. we may assume that $J=\{0,1,2,3\}$ is a set of information positions. Then $\mathcal{F}=\{\{v_0,v_1\},\{v_0,v_2\},\{v_1,v_3\},\{v_2,v_3\}\}$ is an $J$-admissible set of $1$-flats.

\item[(b)] $\mu(1,m)=\frac{(m+1)(2^m-m-4)}{2}$ for $m \geq 4$:\\
Again we may assume that $J=\{0,1,\ldots,m\}$ is a set of information positions (note that $k=m+1$). Then
$\mathcal{F}=\{\{v_j,v_{j^\prime}\}: j=0,\dots,m$, $j^\prime=j+1, \ldots, 2^{m-1}-2\}$ is an $J$-admissible set of $1$-flats and
$\left|\mathcal{F}\right|=\sum_{i=0}^{m}(2^{m-1}-2-i)=(m+1)(2^{m-1}-2)-\frac{m(m+1)}{2}=\frac{(m+1)(2^m-m-4)}{2}$.

\end{enumerate}

\subsection{$\text{RM}(2,4)$}\label{subsect_RM2_4}

We have $\mu(2,4)=7$:\\
Here, $k=11$. Let $J=\{0,1,\ldots,10\}$. Every sum of three of the vectors $v_{11},\ldots,v_{15}$ is contained in $V_J:=\{v_0,\ldots,v_{10}\}$, since otherwise $\{0,\ldots,15\}\setminus J$ would contain $\text{supp}(\chi_U)$ for some $2$-flat $U$ in $\mathbb{Z}_2^{4}$. Since $\chi_U \in \text{RM}(2,4)$, this contradicts the fact that $J$ is a set of information positions (cf.~\ref{nota}~(a)). By the same reason, $v_{11}+ \cdots + v_{15}\in V_J$ and two $3$-sums of $v_{11},\ldots,v_{15}$ are distinct and also not equal to $v_{11} +\cdots + v_{15}$. Therefore, $V_J=\{v_{11} + \cdots + v_{15}, v_g+v_h+v_i, 11 \leq g< h<i \leq 15\}$. Now, in order to have six \mbox{$2$-flats} intersecting pairwise in at most one $v_j \in V_J$ and such that each $v_j \in V_J$ is contained in two of these \mbox{$2$-flats}, either four of these have to consist entirely of vectors in $V_J$ and two have to contain three vectors from $V_J$, or five consist of vectors in $V_J$ and one contains exactly two vectors from $V_J$. Now, by a somewhat tedious calculation (distinguishing the cases whether $v_{11}+ \cdots + v_{15}$ is contained in none, one, or two of the four or five $2$-flats in $V_J$) it can be shown that this is not possible. Hence, $\mu(2,4)>6$. The conclusion $\mu(2,4)=7$ follows from the fact that the following is an $J$-admissible family of $2$-flats where $J$ corresponds to the set $\mathbb{Z}_2^{4}\setminus \{e_1,e_2,e_3,e_4,e_1+e_2+e_3+e_4\}$, $e_1,e_2,e_3,e_4$ being a basis of $\mathbb{Z}_2^{4}$:
\begin{align*}
&\{0,e_2+e_3,e_1+e_2+e_4,e_1+e_3+e_4\},\\
&\{0,e_1+e_4,e_1+e_2+e_3,e_2+e_3+e_4\},\\
&\{e_1,e_1+e_2,e_1+e_3,e_1+e_2+e_3\},\\
&\{e_2,e_1+e_3,e_3+e
_4,e_1+e_2+e_4\},\\
&\{e_2,e_2+e_3,e_2+e_4,e_2+e_3+e_4\},\\
&\{e_4,e_2+e_4,e_1+e_3+e_4,e_1+e_2+e_3+e_4\},\\
&\{e_1+e_2,e_1+e_4,e_2+e_3,e_3+e_4\}.
\end{align*}

\subsection{$\text{RM}(2,5)$}\label{subsect_RM2_5}

All calculations in this section have been performed with \textsf{GAP}~\cite{GAP}. We identify $\mathbb{Z}_2^5$ with $\mathbb{F}_{32}= \mathbb{Z}_{2}[x] / (1+x^2+x^5)$. Then  $\mathbb{Z}_2^5=\{\alpha^0,\alpha^1,\ldots,\alpha^{30},0\}$ where $\alpha^5=\alpha^2+1$. Therefore we let position $j$, $0 \leq j \leq 30$, of a word in $\mathbb{F}_2^{32}$ correspond to $\alpha^j$ and position $31$ to $0$ and consider $\text{RM}(2,5)$ with respect to this ordering. (Note that this is not the lexicographic ordering in $\mathbb{Z}_2^5$; cf. the remark at the beginning of Section~\ref{sect_2step}). Occasionally, we write $v_j$ for $\alpha^j$ and $v_{31}$ for $0$.

Since the determination of $\mu(2,5)$ requires the consideration of all possible sets of information positions for $\text{RM}(2,5)$, we need representatives of the orbits of $\text{AGL}(5,2)=\text{Aut}(\text{RM}(2,5))$ on the $16$-subsets of $\mathbb{Z}_2^5$ that correspond to information positions.

\begin{proposition}\label{prop_sect4}
There are seven orbits of $\text{AGL}(5,2)$ on the $16$-sets of information positions of $\text{RM}(2,5)$. Using lexicographic order in these sets, the first member in each orbit is given in the following list:
\begin{align*}{}
(1)\quad & (0,1,2,3,4,5,6,7,8,9,10,11,12,13,14,15)\\
(2)\quad & (0,1,2,3,4,5,6,7,8,9,10,11,12,13,14,16)\\
(3)\quad & (0,1,2,3,4,5,6,7,8,9,10,11,12,13,15,26)\\
(4)\quad & (0,1,2,3,4,5,6,7,8,9,10,11,12,13,16,20)\\
(5)\quad & (0,1,2,3,4,5,6,7,8,9,10,11,13,15,17,19)\\
(6)\quad & (0,1,2,3,4,5,6,7,8,9,10,11,13,15,19,20)\\
(7)\quad & (0,1,2,3,4,5,6,7,8,9,12,13,15,19,27,31)\\
         & \mbox{The lexicographically first one in this orbit not}\\
	 & \mbox{containing position $31$ is}\\
         & (0,1,2,3,4,5,6,7,8,9,12,13,17,19,20,27).
\end{align*}
\end{proposition}

We have listed in $(7)$ also a representative without position $31$ since puncturing of $\text{RM}(2,5)$ at position $31$ yields a cyclic $[31,16,7]$-code which is also useful in applications and to which Chen's majority-logic procedure and its modification described in the present paper also applies.

In light of the remark at the beginning of Section~\ref{sect_infopos}, we present the generator matrix of $\text{RM}(2,5)$ in systematic form for information positions of type $(1)$ in Table~\ref{Table1b}. The dashed line preceding the last column has been inserted as the first $31$ columns constitute a generator matrix for the punctured cyclic $[31,16,7]$-code.

\begin{table*}[!t] %[!t] for top table* for two-columns floats
\renewcommand{\arraystretch}{1.3}
\caption{Generator matrix of $\text{RM}(2,5)$ in systematic form for information positions of type $(1)$.}
\label{Table1b}
\begin{center}
\[\left(\begin{array}{c@{\quad}c@{\quad}c@{\quad}c@{\quad}c@{\quad}c@{\quad}c@{\quad}c@{\quad}c@{\quad}c@{\quad}c@{\quad}c@{\quad}c@{\quad}c@{\quad}c@{\quad}c@{\quad}c@{\quad}c@{\quad}c@{\quad}c@{\quad}c@{\quad}c@{\quad}c@{\quad}c@{\quad}c@{\quad}c@{\quad}c@{\quad}c@{\quad}c@{\quad}c@{\quad}c@{\quad}|c@{\quad}}
1 & 0 & 0 & 0 & 0 & 0 & 0 & 0 & 0 & 0 & 0 & 0 & 0 & 0 & 0 & 0 & 1 & 1 & 1 & 1 & 0 & 1 & 0 & 1 & 1 & 1 & 1 
& 1 & 0 & 0 & 0 & 1\\
0 & 1 & 0 & 0 & 0 & 0 & 0 & 0 & 0 & 0 & 0 & 0 & 0 & 0 & 0 & 0 & 0 & 1 & 1 & 1 & 1 & 0 & 1 & 0 & 1 & 1 & 1 & 1 
& 1 & 0 & 0 & 1\\
0 & 0 & 1 & 0 & 0 & 0 & 0 & 0 & 0 & 0 & 0 & 0 & 0 & 0 & 0 & 0 & 0 & 0 & 1 & 1 & 1 & 1 & 0 & 1 & 0 & 1 & 1 & 1 & 1 
& 1 & 0 & 1\\ 
0 & 0 & 0 & 1 & 0 & 0 & 0 & 0 & 0 & 0 & 0 & 0 & 0 & 0 & 0 & 0 & 0 & 0 & 0 & 1 & 1 & 1 & 1 & 0 & 1 & 0 & 1 & 1 & 1 & 1 
& 1 & 1\\ 
0 & 0 & 0 & 0 & 1 & 0 & 0 & 0 & 0 & 0 & 0 & 0 & 0 & 0 & 0 & 0 & 1 & 1 & 1 & 1 & 1 & 0 & 1 & 0 & 1 & 0 & 1 &  
0 & 1 & 1 & 1 & 0\\
0 & 0 & 0 & 0 & 0 & 1 & 0 & 0 & 0 & 0 & 0 & 0 & 0 & 0 & 0 & 0 & 1 & 0 & 0 & 0 & 1 & 0 & 0 & 0 & 1 & 0 & 1 & 
0 & 0 & 1 & 1 & 1\\
0 & 0 & 0 & 0 & 0 & 0 & 1 & 0 & 0 & 0 & 0 & 0 & 0 & 0 & 0 & 0 & 1 & 0 & 1 & 1 & 0 & 0 & 0 & 1 & 1 & 0 & 1 & 
0 & 0 & 0 & 1 & 0\\
0 & 0 & 0 & 0 & 0 & 0 & 0 & 1 & 0 & 0 & 0 & 0 & 0 & 0 & 0 & 0 & 1 & 0 & 1 & 0 & 1 & 1 & 0 & 1 & 0 & 0 & 1 & 
0 & 0 & 0 & 0 & 1\\ 
0 & 0 & 0 & 0 & 0 & 0 & 0 & 0 & 1 & 0 & 0 & 0 & 0 & 0 & 0 & 0 & 0 & 1 & 0 & 1 & 0 & 1 & 1 & 0 & 1 & 0 & 0 & 
1 & 0 & 0 & 0 & 1\\
0 & 0 & 0 & 0 & 0 & 0 & 0 & 0 & 0 & 1 & 0 & 0 & 0 & 0 & 0 & 0 & 0 & 0 & 1 & 0 & 1 & 0 & 1 & 1 & 0 & 1 & 0 & 0 & 
1 & 0 & 0 & 1\\
0 & 0 & 0 & 0 & 0 & 0 & 0 & 0 & 0 & 0 & 1 & 0 & 0 & 0 & 0 & 0 & 0 & 0 & 0 & 1 & 0 & 1 & 0 & 1 & 1 & 0 & 1 & 0 & 0 & 
1 & 0 & 1\\
0 & 0 & 0 & 0 & 0 & 0 & 0 & 0 & 0 & 0 & 0 & 1 & 0 & 0 & 0 & 0 & 0 & 0 & 0 & 0 & 1 & 0 & 1 & 0 & 1 & 1 & 0 & 1 & 0 & 0 & 
1 & 1\\
0 & 0 & 0 & 0 & 0 & 0 & 0 & 0 & 0 & 0 & 0 & 0 & 1 & 0 & 0 & 0 & 1 & 1 & 1 & 1 & 0 & 0 & 0 & 0 & 1 & 0 & 0 & 1 
& 1 & 0 & 0 & 0\\
0 & 0 & 0 & 0 & 0 & 0 & 0 & 0 & 0 & 0 & 0 & 0 & 0 & 1 & 0 & 0 & 0 & 1 & 1 & 1 & 1 & 0 & 0 & 0 & 0 & 1 & 0 & 0 & 1 
& 1 & 0 & 0\\ 
0 & 0 & 0 & 0 & 0 & 0 & 0 & 0 & 0 & 0 & 0 & 0 & 0 & 0 & 1 & 0 & 0 & 0 & 1 & 1 & 1 & 1 & 0 & 0 & 0 & 0 & 1 & 0 & 0 & 1 
& 1 & 0\\
0 & 0 & 0 & 0 & 0 & 0 & 0 & 0 & 0 & 0 & 0 & 0 & 0 & 0 & 0 & 1 & 1 & 1 & 1 & 0 & 1 & 0 & 1 & 1 & 1 & 1 & 1 & 0 
& 0 & 0 & 1 & 1 
\end{array}\right)\]
\end{center}
\end{table*}

It might be of interest to list some properties of the seven orbits of information positions. It turns out that they can be distinguished already by one parameter $a$. We define this and some more as follows:
Let $\mathcal{O}$ be one of the seven orbits of $16$-sets of information positions given in Proposition~\ref{prop_sect4}, and let $J \in \mathcal{O}$. Let $V_J$ be the corresponding set of vectors in $\mathbb{Z}_2^5$. Define

\begin{itemize}
\item $l=$ length of $\mathcal{O}$,
\item $a=$ number of $6$-subsets in $V_J$ consisting of affinely independent vectors,
\item $n_i=$ number of $2$-flats of $\mathbb{Z}_2^5$ intersecting $V_J$ in exactly $i$ elements ($i=0,1,2,3,4$),
\item $c=$ maximal number of $2$-flats contained in $V_J$ which intersect pairwise in at most one vector
(the abbreviation $c$ stands for `clique' since $c$ is just the maximal size of a clique of the graph whose vertices are the $2$-flats in $V_J$ and where two vertices are joined by an edge if their intersection has size at most $1$),
\item $n_{\text{max}}=$ number of cliques of maximal size (in the above sense).
\end{itemize}

\begin{proposition}\label{prop_sect4_2}
We obtain the following properties, as displayed in Table~\ref{Table2}.
\end{proposition}

\begin{IEEEproof}
By computer calculations.
\end{IEEEproof}

\begin{table*}[!t]
\renewcommand{\arraystretch}{1.3}
\caption[]{Some properties of the seven orbits of $\text{AGL}(5,2)$ on the $16$-sets of\\ information positions of $\text{RM}(2,5)$.}\label{Table2}

\begin{center}
\begin{tabular}{|c|c|c|c|c|c|c|c|c|c|}
  \hline
  \mbox{Type of orbit} & $l$ & $a$ & $n_0$ & $n_1$ & $n_2$ & $n_3$ & $n_4$ & $c$ & $n_{\text{max}}$\\
  \hline \hline
1 & 31.997.952 & 4{.}051 & 60 & 320 & 480 & 320 & 60 & 12 & 105\\
2 & 79.994.880 & 4{.}004 & 61 & 316 & 486 & 316 & 61 & 13 & 8\\
3 & 53.329.920 & 3{.}959 & 62 & 312 & 492 & 312 & 62 & 11 & 922\\
4 &  5.332.992 & 4{.}052 & 60 & 320 & 480 & 320 & 60 & 12 & 20\\
5 & 19.998.720 & 3{.}912 & 63 & 308 & 498 & 308 & 63 & 12 & 40\\
6 &  6.666.240 & 4{.}000 & 61 & 316 & 486 & 316 & 61 & 15 & 6\\
7 &    444.416 & 3{.}816 & 65 & 300 & 510 & 300 & 65 & 15 & 6\\
\hline
\end{tabular}
\end{center}
\end{table*}

We turn now to the determination of $\mu(2,5)$. In this case, this means to find the minimal number of $2$-flats in $\mathbb{Z}_2^5$ such that for a suitable set $J$ of information positions and each $j \in J$ there exist six among these $2$-flats pairwise intersecting in $v_j$. We obtain the following result.

\begin{table*}[!t]
\renewcommand{\arraystretch}{1.3}
\caption[]{Numbers of large cliques of $2$-flats with four information positions\\ for the seven types of information positions.}\label{Table3}

\begin{center}
\begin{tabular}{|c||c|c|c|c|c|c|c|}
  \hline
  \mbox{Type} & $(1)$ & $(2)$ & $(3)$ & $(4)$ & $(5)$ & $(6)$ & $(7)$\\
  \hline \hline
\mbox{Number of cliques of size} &  &  &  & &  &  &\\
9 & 194{.}670 & 182{.}940 & 111{.}966 & 192{.}880 & 105{.}516 & 354{.}830 & 198{.}170\\
10 & 36{.}771 & 37{.}806 & 16{.}200 & 36{.}928 & 18{.}584 & 125{.}342 & 63{.}658\\
11 & 3{.}300 & 4{.}512 & 922 & 2{.}530 & 1{.}888 & 36{.}762 & 17{.}870\\
12 & 105 & 292 & -- & 20 & 40 & 8{.}490 & 4{.}170\\
13 & -- & 8 & -- & -- & -- & 1{.}398 & 750\\
14 & -- & -- & -- & -- & -- &  138 & 90\\
15 & -- & -- & -- & -- & -- & 6 & 6\\
\hline
\end{tabular}
\end{center}
\end{table*}

\begin{theorem}\label{minnumb_gates}
$\mu(2,5) =30$.
\end{theorem}

\begin{IEEEproof}
For the lower bound, note that by Table~\ref{Table1}, $\mu(2,5) \geq 28$. Suppose that actually $\mu(2,5) = 28$, and let $J$ denote a set of information positions such that $\text{min}_J(2,5)=28$. As indicated in Table~\ref{Table1}, $\mu(2,5) = 28$ means that there is a choice of $28$ $2$-flats which corresponds to the minimum solution of the ILP in Proposition~\ref{prop3.4}. It is easy to see that there is only one admissible solution of the ILP yielding the minimum, namely $x_4=12$ and $x_3=16$. Recall that $x_i$ denotes the number of $2$-flats that are used at exactly $i$ positions of $J$. Using the uniqueness of this solution some simple combinatorial considerations show that among those $28$ $2$-flats any two have an intersection of size at most $1$. (This means that there do not exist $2$-flats (here denoted by the positions instead of the corresponding vectors) $\{j_1,j_2,j_3,j_4\}$ and $\{j_1,j_2,j_5,j_6\}$, $j_i \in J$, which form a `cross-junction' with respect to $
 j_1,j_2$; by this we mean that the first one is used at positions $j_1,j_3,j_4$ but not at $j_2$ and the second one at positions $j_2,j_5,j_6$ but not at $j_1$.) Hence, these $28$ $2$-flats form a clique in the sense used before, containing at least $12$ $2$-flats consisting entirely of vectors corresponding to information positions. By Table~\ref{Table2}, the maximal size of a clique of the latter type of $2$-flats is $15$, realized for $J$ of type $(6)$ or $(7)$. So we have to consider such cliques of size $12$--$15$. Fortunately, their number is not too large.  For the seven types of information positions these numbers are given in the last four rows of Table~\ref{Table3}. For all these cases it can be verified by computer that they cannot be completed to cliques of size $28$ with $2$-flats containing exactly three vectors corresponding to information positions. This shows that $\mu(2,5) \geq 29$.

Now, assume that $\mu(2,5) = 29$. There are exactly $18$ solutions $(x_4,x_3,x_2,x_1)$ of  (\ref{E4})--(\ref{E6}) (with $r=2,m=5$) satisfying $\sum_{i=1}^4x_i=29$. 
This number can be reduced to $12$ if we consider only those solutions for which we have equality in (\ref{E5}):
\begin{equation}\label{E5prime}
x_1 + 2 x_2 + 3 x_3+4 x_4=96.\tag{III.4$^\prime$}
\end{equation}
This is possible by deactivating suitably many information positions in $2$-flats of a solution for which  (\ref{E5}) is a strict inequality. The validity of  (\ref{E4}) and  (\ref{E6}) is not affected by this. In all of these solutions, $9 \leq x_4 \leq 15$. Since there are sets of information positions where the maximum number $c$ of cliques is $15$ (cf.~Table~\ref{Table2}), none of these possibilities can be ruled out immediately. Moreover, as shown in Table~\ref{Table3}, the number of different cliques of size $9$, $10$ or $11$ is prohibitively large for a brute force attempt. Therefore, a refined analysis of this case is needed for a feasible computational approach.
Given a set $J$ of information positions with $\text{min}_J(2,5)=29$ and a corresponding solution $\mathcal{F}$ of all $2$-flats with parameters $(x_4,x_3,x_2,x_1)$, $\sum_{i=1}^4x_i=\left| \mathcal{F} \right| =29$, let $\mathcal{F}_{is}$, $1 \leq i \leq s \leq 4$, denote the set of those $2$-flats in $\mathcal{F}$ that contain exactly $s$ vectors corresponding to positions in $J$, $i$ of which are active. Setting $x_{is}=\left|\mathcal{F}_{is}\right|$, we have $\sum_{s=i}^4x_{is}=x_i$ for $i=1,\ldots,4$. We first note that
\begin{equation}\label{NE1}
\sum_{1\leq i\leq s \leq 4} i(s-1)x_{is} \leq 240.
\end{equation}
This follows from the fact that for $U \in \mathcal{F}_{is}$ there exist \mbox{$i(s-1)$} pairs $(j,j^\prime) \in J \times J$ with $j$ active in $U$ and $v_{j^\prime} \in U$, $j \neq j^\prime$. Since any two members in $\mathcal{F}$ which are used for the same position have intersection size $1$, the sum on the left hand side of~(\ref{NE1}) is bounded above by the number of all pairs $(j,j^\prime) \in J \times J$, $j \neq j^\prime$, which is $240$. 

It is now straightforward to verify that (\ref{E4}), (\ref{E5prime}\renewcommand{\theequation}{{III.4}$^\prime$}), (\ref{E6}) and (\ref{NE1}) imply $x_{34} \leq 4$.
With this information we strengthen (\ref{E6}) which states that $\sum_{i=2}^4 \binom{i}{2} x_i \leq 120$. We recall that the left hand side counts the number of $2$-sets of active positions in $U$, $U$ ranging over $\mathcal{F}$; the right hand side is just the number of all $2$-subsets of $J$.
Let $\{j,j^\prime\}$ be a $2$-subset of $J$ and suppose there exists $U \in \mathcal{F}_{34}$ such that $j$ is active in $U$ and $j^\prime$ not but $v_{j^\prime} \in U$ or vice versa. Such a set is not counted in the left hand side of (\ref{E6}) and can only be obtained by at most two $U,U^\prime \in \mathcal{F}_{34}$ which then form a cross-junction with respect to $j,j^\prime$ in the sense described above. Hence if $N$ is the number of of cross-junctions between members in $\mathcal{F}_{34}$, it follows that the left hand side of (\ref{E6}) is bounded above by $120-(3x_{34}-N)$. Now for $x_{34} \leq 4$ it is easy to see that $N \leq x_{34}-1$ if $N \neq 0$. 
Hence we obtain 
\begin{equation}\label{NE2}
\sum_{2 \leq i \leq s \leq 4} \binom{i}{2} x_{is} + 2 x_{34} \leq  \begin{cases}
120, & \text{if} \;\,x_{34} =0\\
119, & \text{if} \;\, 1 \leq x_{34} \leq 4.
\end{cases}\tag{IV.2}
\end{equation}
From this and (\ref{E4}), (\ref{E5prime}) it follows easily that
\[x_{34} \leq 2 \quad \text{and} \quad 8 \leq x_{33} \leq 20.\]
These restrictions together with some stopping criteria derived from (\ref{E4}), (\ref{E5prime}), (\ref{NE1}) and (\ref{NE2}) give rise to a backtracking algorithm which efficiently rules out the case $\mu(2,5) = 29$.

To complete the proof of this theorem we have to present an $J$-admissible family of $30$ $2$-flats for some set $J$ of information positions. The example we give belongs to the representative $J$ of type $(1)$ exhibited in Proposition~\ref{prop_sect4}. This example has $x_4=9$, $x_3=18$, $x_2=3$ (the corresponding $2$-flats are displayed below in this order) and all $2$-flats are used at each of their information positions. For the sake of simplicity we use the numbers of the positions instead of the corresponding vectors to describe these $2$-flats:
\begin{align*}
&\{0,1,8,12\},\,    \{0,4,5,7\},\,    \{1,6,7,13\},\,  \{1,9,11,15\},\\
&\{2,4,9,12\},\,  \{2,6,10,15\},\,   \{2,7,8,14\},\, \{3,5,10,13\},\\
&\{4,6,11,14\},\\
&\{0,2,13,25\},\,  \{0,3,9,17\},\,   \{0,10,11,26\},\,  \{0,14,15,18\},\\
&\{1,3,14,26\},\,    \{1,4,10,18\},\,   \{2,5,11,19\},\, \{3,4,8,22\},\\
&\{3,6,12,20\},\,   \{3,7,11,16\},\,    \{4,13,15,17\},\, \{5,6,9,22\},\\
&\{5,8,15,26\},\,   \{5,12,14,28\},\, \{7,10,12,27\},\,  \{8,9,10,19\},\\
&\{9,13,14,16\},\,   \{11,12,13,22\},\\
&\{1,2,17,22\},\, \{6,8,24,25\},\,  \{7,15,25,30\}.
\end{align*}

\end{IEEEproof}

\begin{remark}\label{final}

\begin{enumerate}

\item[(a)] The example with $30$ $2$-flats given in Theorem~\ref{minnumb_gates} has the property that none of its $2$-flats contains position $31$ (in fact also not $21$, $23$, $29$). Note that in Chen's decoding procedure (\ref{chen}) for each of these $2$-flats six $3$-flats containing it are needed to compute the bilinear forms in Step~1. By Lemma~\ref{lem_flats}, there are actually seven such $3$-flats and it is easy to see that exactly one of them contains $0\in \mathbb{Z}_2^5$ (corresponding to position $31$). Choosing always the six $3$-flats that do not contain $0$ shows that the $30$ $2$-flats above can also be used for the two-step majority-logic decoding at the information positions of type $(1)$ of the cyclic $[31,16,7]$-code obtained by puncturing $\text{RM}(2,5)$ at position $31$.

\item[(b)] We have also found solutions with $30$ $2$-flats for information positions of all other types $(2)$--$(7)$. Moreover, there are solutions with $30$ $2$-flats and different values of  $x_1,\ldots, x_4$ than those in the example presented in the proof of Theorem~\ref{minnumb_gates}.

\end{enumerate}

\end{remark}

\section{Complexity comparison with Chen's algorithm}\label{comparison}

\begin{table}[!t] %[!t] for top table, * for two-columns floats
\renewcommand{\arraystretch}{1.3}
\caption{Some complexity bounds for short-length $\text{RM}(r,m)$.}\label{Table_comparison}

\begin{center}
\begin{tabular}{|c|c||c|c|}
  \hline
   \multirow{2}{*}{$r$} & \multirow{2}{*}{$m$} & Chen's & Improved \\
      &  & majority-logic decoding & majority-logic decoding\\
  \hline \hline
  1 &  3           &  $16$ &  $8$ \\
  1 &  4           &   $64$    & $25$ \\  
  2 & 4            &    $24$   & $18$\\
  2 & 5           &     $80$  & $46$\\
  2 & 6          &    $288$  & $\leq 231$\\
  2 & 7         & $1{.}088$ & $\leq 878$ \\
  3 & 6         &   $122$  & $\leq 90$\\
  3 & 7         &    $352$  & $\leq 265$\\
  3 & 8         &  $1{.}216$ & $\leq 1{.}013$\\ 
  4 & 8         &  $480$ & $\leq 359$\\  
  4 & 9         &  $1{.}472$ & $\leq 1{.}214$\\
  4 & 10       &  $4{.}992$ & $\leq 4{.}340$\\    \hline 
\end{tabular}
\end{center}
\end{table}

We compare in this section our improved majority-logic decoding method with Chen's decoding algorithm for various Reed--Muller codes up to length $2^{10}$. As shown in Table~\ref{Table_comparison}, our decoding procedure yields a significant gain over Chen's procedure. The minimal number of gates needed for Chen's majority-logic decoding method are indicated in the first column. The second column contains the exact number of majority gates, if known, needed for the improved decoding, or upper bounds using Proposition~\ref{prop_sect3}.

\section{Conclusion}\label{conclusion}

Short-length Reed--Muller codes under majority-logic decoding are of particular importance for efficient hardware implementations in real-time and embedded systems. In this paper, the two-step majority-logic decoding method for binary Reed--Muller codes $\text{RM}(r,m)$, $r \leq m/2$, as introduced by Chen has been investigated. If errors at all positions of a word are to be corrected, assuming that at most $2^{m-r-1}-1$ errors occurred, $2^{m-r}(2^{m-r}-2)$ majority gates for the first step and $2^m$ majority gates for the second step are needed. If, however, under systematic encoding only errors at the chosen information positions are to be corrected, the number of majority gates can be reduced significantly. Some general results that are particularly good for short codes have been presented and, with its importance in applications as a $3$-error-correcting, self-dual code, the smallest non-trivial example, $\text{RM}(2,5)$ of dimension 16 and length 32, has been investigat
 ed in detail. To correct errors at suitably chosen 16 information positions, the minimal number of majority gates for the second step is clearly 16 instead of 32. It has been shown that the minimal number of majority gates for the first step is 30. This is to be compared with 48 gates needed for the first step if all positions are to be corrected. We have furthermore compared the decoding complexity of our procedure with that of Chen's decoding algorithm for various Reed--Muller codes up to length $2^{10}$.

\appendix[Proof of Proposition~\ref{prop_sect3}]

We first prove the following lemma, which holds without our usual restriction $r \leq m/2$.

\begin{lemma}\label{lem_sect3}
Let $e_1,\ldots,e_m$ be a basis of $\mathbb{Z}_2^m$.

\begin{enumerate}

\item[(a)] Let $J$ be the set of indices from $\{0,\ldots,2^m-1\}$ with the property that $j\in J$ if and only if $v_j$ is a linear combination of at least $m-r$ of the vectors $e_1,\ldots,e_m$ (i.e. if $\text{wt}(v_j) \geq m-r$ with respect to the basis $e_1,\ldots,e_m$). Then $J \in I(r,m)$.

\item[(b)] Let $ 0 \leq i \leq r$. Then there exists a family $\mathcal{R}_i$ of $r$-flats in $\mathbb{Z}_2^m$ with
\[\left|\mathcal{R}_i\right| =\binom{m-i}{r-i} + \bigg\lceil\frac{1}{r+1} \cdot \sum_{s=0}^{i-1} 2^s \cdot \binom{m-1-s}{r-s} \bigg\rceil\]
such that every vector in $\mathbb{Z}_2^m$ of weight $\geq m-r$ with respect to $e_1,\ldots,e_m$ is contained in at least one $r$-flat from $\mathcal{R}_i$. Among those $\mathcal{R}_i$, the one for $i=\lceil \log_2(r+1)\rceil$ has minimal size.
\end{enumerate}

\end{lemma}

\begin{IEEEproof}
(a): For any subset $I=\{i_1,\ldots,i_l\} \subseteq\{1,\ldots,m\}$ of size $l\geq m-r$, let  $U_I=\langle e_{i_1},\ldots,e_{i_l}\rangle$. Clearly, $\chi_{U_I} \in \text{RM}(r,m)$. Since $e_{i_1} + \cdots +  e_{i_l}$ is the unique vector of maximal weight in $U_I$, it is clear that the $\chi_{U_I}$ are linearly independent in $\mathbb{Z}_2^{2^m}$ and hence form a basis of  $\text{RM}(r,m)$. With respect to a suitable ordering of the subsets $I$ considered above, the row vectors $\chi_{U_I}$ constitute a $k \times 2^m$-matrix $M$ ($k=\sum_{i=m-r}^{m} \binom{m}{i}=\sum_{i=0}^{r} \binom{m}{i}$) such that the $k \times k$-matrix  obtained by using the columns of $M$ at positions corresponding to vectors in  $\mathbb{Z}_2^m$ of weight $\geq m-r$ with  respect to $e_1,\ldots,e_m$ is a triangular matrix with entries $1$ along the diagonal. The assertion follows.

(b): Let $ 0 \leq i \leq r$. Set $e=e_1 + \cdots + e_m$ and define
\begin{align*}{}
\mathcal{S}_i=\left\{ e+U : 
\begin{array}{l}
U\text{ is an $r$-dimensional subspace of }\mathbb{Z}_2^m\\
\text{containing } e_1,\ldots,e_i 
\end{array}
\right\}.
\end{align*}

Clearly, $\left|\mathcal{S}_i\right|=\binom{m-i}{r-i}$.
It is easily checked that there are exactly $\sum_{s=0}^{i-1} 2^s \cdot \binom{m-1-s}{r-s}$ vectors of weight $\geq m-r$ with respect to $e_1,\ldots,e_m$ in $\mathbb{Z}_2^{m}$ which are not contained in any member of $\mathcal{S}_i$. Any $r+1$ of them are contained in an $r$-flat of $\mathbb{Z}_2^{m}$. Hence there is a family $\mathcal{T}_i$ of $r$-flats with $\left|\mathcal{T}_i\right| = \left\lceil\frac{1}{r+1} \cdot \sum_{s=0}^{i-1} 2^s \cdot \binom{m-1-s}{r-s} \right\rceil$ containing all those vectors.
Putting $\mathcal{R}_i=\mathcal{S}_i \cup \mathcal{T}_i$ yields a family of $r$-flats in $\mathbb{Z}_2^{m}$ containing all vectors of weight $\geq m-r$.
That $\left|\mathcal{R}_i\right|$ is minimal for $i=\lceil \log_2(r+1)\rceil$ follows from a straightforward calculation.
\end{IEEEproof}

We are now ready to prove Proposition~\ref{prop_sect3}:
\balance
\begin{IEEEproof}
(a): We note that $\text{min}_J(r,m) \leq k \cdot (2^{m-r}-2)$ holds for any set $J$ of information positions, since for every position the required $r$-flats (which exist by Lemma~\ref{lem_flats}~(b)) are counted separately.
In the following, we take $J$ to be the set of positions corresponding to vectors of weight $\geq m-r$ as in Lemma~\ref{lem_sect3}~(a).
Let $j \in J$, and let $U_1,\ldots,U_{2^{m-r}-2}$ be $r$-flats in $\mathbb{Z}_2^{m}$ such that $U_s \cap U_{s^\prime} =\{v_j\}$ for all $s,s^\prime \in \{1,\ldots,2^{m-r}-2\}$, $s \neq s^\prime$. Let $U$ be an arbitrary $r$-flat containing $v_j$. It follows from the transitivity of the stabilizer of $v_j$ in $\text{AGL}(m,2)=\text{Aut}(\text{RM}(r,m))$~\cite[Ch.\,17, Thm.\,4.4]{HB98} on the set of $r$-flats containing $v_j$ that there exists $\rho \in \text{AGL}(m,2)$ with $U_1\rho=U$, $v_j\rho=v_j$. Then $U_1\rho=U, U_2\rho,\ldots, U_{2^{m-r}-2} \rho$ is also a family of $r$-flats with the required properties for position $j$. If we take the family $\mathcal{R}=\mathcal{R}_b$ of $r$-flats constructed in Lemma~\ref{lem_sect3}~(b), the argument above shows that for any position $j\in J$ there are $2^ {m-r}-2$ $r$-flats for position $j$ where one of them can be chosen from $\mathcal{R}$. Now the assertion follows from Lemma~\ref{lem_sect3}~(b).

(b): We refer to the construction in the proof of Proposition~\ref{prop_inters} and show that for a suitable choice of the $U_p,W_p$, $p=1,\ldots , 2^{m-r}-2$, there are at least $\left\lfloor\frac{m}{r} \right\rfloor \cdot \sum_{s=0}^{m-2r-1} \binom{m-r}{s}$ $r$-flats that do not contain any vector $v_j$ for $j\in J$, $J$ the set of information positions as in Lemma~\ref{lem_sect3}~(a). These can then be omitted from the $2^{m-r}(2^{m-r}-2)$ $r$-flats of Proposition~\ref{prop_inters} if only errors at positions of $J$ are to be corrected. Let $e_1, \ldots, e_m$ be the basis of $\mathbb{Z}_2^{m}$ such that every $v_j$, $j \in J$, has weight $\geq m-r$ with respect to $e_1,\ldots,e_m$. Let
$l=\left\lfloor\frac{m}{r} \right\rfloor$. There are $2^{m-r}-2$ subspaces $U_p$ of dimension $r$ pairwise intersecting in $0$. By applying a suitable transformation in $\text{GL}(m,2)$ we may assume that $U_p=\langle e_{r\cdot p-(r-1)},\ldots,e_{r \cdot p}\rangle$, $p=1,\ldots,l$. We let $W_p=\langle e_{i}: i \in \{1,\ldots,m\} \setminus \{r\cdot p -(r-1), \ldots, r\cdot p\}\rangle$; hence $\mathbb{Z}_2^{m}=U_p\oplus W_p$ for $p=1,\ldots,l$. $W_p$ contains $\sum_{s=0}^{m-2r-1} \binom{m-r}{s}$ vectors of weight less than $m-2r$. If $w$ is such a vector, then $w+U_p$ is an $r$-flat containing only vectors of weight less than $m-r$. Totally we obtain this way $l \cdot \sum_{s=0}^{m-2r-1} \binom{m-r}{s}$ $r$-flats as members of the family constructed in Proposition~\ref{prop_inters} that are not needed to correct errors at positions in $J$.
\end{IEEEproof}

\section*{Acknowledgment}

The authors thank Siegbert Steinlechner from Robert Bosch GmbH for bringing the problem considered in this paper to their attention and for sharing his ideas and first results about the case of $\text{RM}(2,5)$. The authors also thank the anonymous referees for their thorough reviews and valuable comments that helped improving the presentation of the paper.

\end{document}